\documentclass[a4paper,twocolumn,showpacs,prl,superscriptaddress,floatfix]{revtex4}
\usepackage[german,swedish,english]{babel}
\usepackage[dvips]{graphicx}
\usepackage{amssymb}
\usepackage{amsmath}
\usepackage{color}
\newcommand{\bm}{\mathbf}

\begin{document}

\date{\today}

\title{Exploring the fragile antiferromagnetic superconducting phase in CeCoIn$_5$}

\author{E. Blackburn}
\affiliation{School of Physics and Astronomy, University of Birmingham, B15 2TT, United Kingdom\\}
\author{P. Das}
\affiliation{Department of Physics, University of Notre Dame, Notre Dame, IN 46556, USA\\}
\author{M. R. Eskildsen}
\affiliation{Department of Physics, University of Notre Dame, Notre Dame, IN 46556, USA\\}
\author{E. M. Forgan}
\affiliation{School of Physics and Astronomy, University of Birmingham, B15 2TT, United Kingdom\\}
\author{M. Laver}
\affiliation{Laboratory for Neutron Scattering, Paul Scherrer Institut, CH-5232 Villigen, Switzerland\\}
\affiliation{Materials Research Division, Ris{\o} DTU, Technical University of Denmark, DK-4000 Roskilde, Denmark\\}
\affiliation{Nano-Science Center, Niels Bohr Institute, University of Copenhagen, DK-2100 K{\o}benhavn, Denmark\\}
\author{C. Niedermayer}
\affiliation{Laboratory for Neutron Scattering, Paul Scherrer Institut, CH-5232 Villigen, Switzerland\\}
\author{C. Petrovic}
\affiliation{Brookhaven National Laboratory, Upton, NY 11973, USA\\}
\author{J. S. White}
\affiliation{School of Physics and Astronomy, University of Birmingham, B15 2TT, United Kingdom\\}
\affiliation{Laboratory for Neutron Scattering, Paul Scherrer Institut, CH-5232 Villigen, Switzerland\\}

\begin{abstract}
CeCoIn$_5$ is a heavy fermion Type-II superconductor which exhibits clear indications of Pauli-limited superconductivity. A variety of measurements give evidence for a transition at high magnetic fields inside the superconducting state, when the field is applied either parallel to or perpendicular to $\bm{c}$ axis.  When the field is perpendicular to the $\bm{c}$ axis, antiferromagnetic order is observed on the high-field side of the transition, with a magnetic wavevector of ($q$ $q$ 0.5), where $q$ = 0.44 reciprocal lattice units.  We show that this order remains as the magnetic field is rotated out of the basal plane, but the associated moment eventually disappears above 17$^{\circ}$, indicating that the anomalies seen with the field parallel to the $\bm{c}$ axis are not related to this magnetic order. We discuss the implications of this finding.
\end{abstract}

\pacs{74.70.Tx,75.30.Kz}

\maketitle

CeCoIn$_5$ is one of the more prominent members of the Ce family of heavy fermion compounds.  It is an ambient pressure superconductor with a zero field $T_c$ of 2.3~K, the highest of any heavy fermion material \cite{PPHM+01}.  It is an extremely clean high-$\kappa$ superconductor, with an electronic mean free path $\sim$ 1 $\mu$m \cite{OSGS+02,KNIM+05}.  It is a strongly Type-II superconductor, and there is clear evidence that the superconductivity is Pauli-limited in strong fields \cite{IYMS+01}.

The material is anisotropic; it has the tetragonal HoCoGa$_5$ structure, with $a$ = 4.602~\AA~ and $c$ = 7.545~\AA~ at 2 K (space group $P4/mmm$).  If the field is applied in the basal plane,the zero-temperature upper critical field $\mu_0 H_{c2} = 11.6$ T, but when the field is parallel to the $\bm{c}$ axis, $\mu_0  H_{c2}$ = 4.95~T.  However, in both cases, it enters the Pauli-limited region close to $H_{c2}$ \cite{KSOF+06}.  A variety of measurements at high fields have found evidence for a transition inside the superconducting mixed state for both field orientations.  These measurements include specific heat \cite{RFMH+03,BMCPS03}, ultrasound studies of the transverse shear velocity \cite{WKIS+04}, thermal conductivity \cite{CBMC+04}, NMR \cite{KSOF+06}, and magnetostriction \cite{CMMP+07}. This transition, existing in a high-field, low-temperature region bordering $H_{c2}$ has been considered as a candidate for a Fulde-Ferrell-Larkin-Ovchinnikov (FFLO) state \cite{FF,LO}. In this state, the Zeeman splitting of the energies of electronic quasiparticles is expected to lead to an additional spatial variation of the superconducting order parameter, for instance~\cite{LO} with nodal planes perpendicular to the flux lines.

\begin{figure}[bth]
\begin{center}
\includegraphics[width=0.39\textwidth]{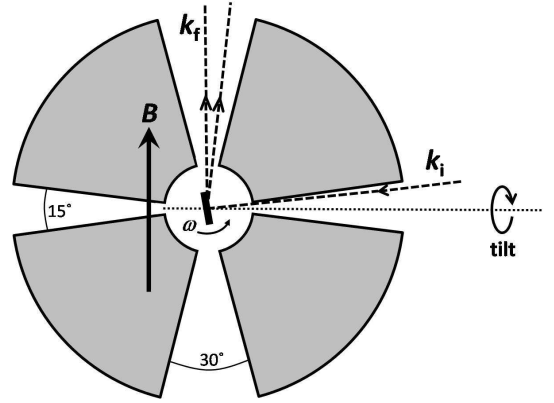}
\end{center}
\caption{A sketch of the geometrical constraints imposed by the cryomagnet.  $\bm{k}_i$ is the incident neutron wavevector and $\bm{k}_f$ is the outgoing neutron wavevector.  For this sample geometry, absorption is minimised if the outgoing (rather than the incoming) beam is nearly parallel to the sample plate.   The circular arrow indicates the direction of tilt required to rotate the magnetic reflections into the scattering plane.}
\label{fig:magnet}
\end{figure}

\begin{figure}[bth]
\begin{center}
\includegraphics[width=0.39\textwidth]{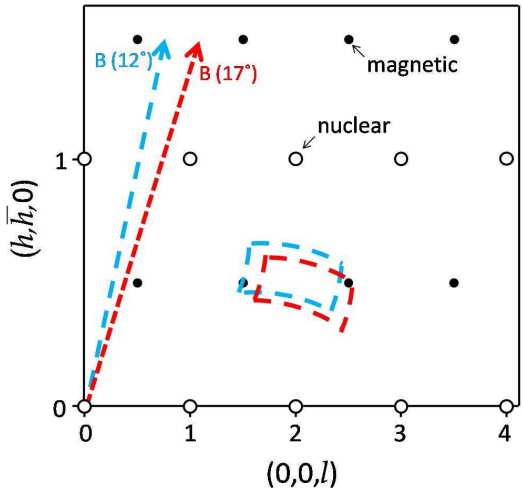}
\end{center}
\caption{(color online) The scattering plane with the cryomagnet level.  The thick dashed lines show the two field orientations examined in this paper.  The marked areas are the reciprocal space accessible with a wavevector of 1.4 \AA $^{-1}$ (the wavevector used to gather the data presented here).  The open circles represent nuclear reflections and the closed points represent magnetic reflections, although the magnet had to be tilted to bring them into the scattering plane (see main text).  For measurements of the nuclear reflections, the direction of the field is irrelevant, and so the sample could be rotated freely to bring them into the window of accessible reciprocal space.}
\label{fig:reciprocal_space}
\end{figure}

Kenzelmann {\it et al.}~\cite{KSNS+08,KGEG+10} have shown that when a magnetic field is applied in the basal plane there is incommensurate antiferromagnetic order inside the high-field phase.  Magnetic Bragg reflections are seen with a wavevector $\pm$($q$ $q$ 0.5) relative to a nuclear peak where $q$ = 0.44 reciprocal lattice units (rlu) when the field is applied parallel to [1 $\bar{1}$ 0].  By symmetry, the wavevector is ($q$ -$q$ 0.5) when the field is parallel to [1 1 0].  The magnetism is carried by the Ce ions, and was found to have the same value of 0.15 $\pm$ 0.05 $\mu_B$/Ce ion and the same ordering vector for two different field directions in the basal plane.  This antiferromagnetic order disappears sharply with increasing field as the superconductivity is destroyed at $ H_{c2}$.  If the magnetic field is applied along [1 0 0], the two wavevectors are degenerate and the sample is expected to have domains of both \cite{KGEG+10}.  An alternative scenario of double-$\bm{q}$ magnetic ordering is ruled out by the observation of a characteristic single-$\bm{q}$ NMR lineshape \cite{CYUG10}. 

We have carried out neutron diffraction measurements on a sample of CeCoIn$_5$, to investigate changes in the antiferromagnetic order when the applied field is rotated out of the basal plane. Our scattering geometry also allows the coherence of the order along the field direction to be investigated. The measurements reported here cast further light on the relationship between the magnetic and superconducting order parameters in CeCoIn$_5$, which remains a matter of debate.

To achieve this geometry, we required a horizontal magnetic field, so that it could be in the scattering plane, and of sufficiently large magnitude to probe the region of interest. For this reason, our experiment could only be carried out at the RITA-II instrument at the SINQ Facility (Paul Scherrer Institut), using the 11 T split-pair horizontal field cryomagnet available there, equipped with a dilution refrigerator (DR) insert.  This magnet has two neutron access ports with an acceptance angle of $\pm$ 15$^{\circ}$, parallel to the field, and  two with an acceptance angle of $\pm$ 7.5$^{\circ}$, perpendicular to the field (Figure \ref{fig:magnet}). The magnet could be rotated about the vertical axis and tilted a few degrees about a horizontal axis.  The DR could be rotated independently about the vertical magnet axis to alter the orientation of the sample relative to the field ($\omega$). However, the limited neutron access (inherent in any high field magnet) imposed the constraint of $2 \theta$ = 90$^{\circ}$ $\pm$ 22.5$^{\circ}$ scattering, severely limiting the accessible region of reciprocal space (see Figure \ref{fig:reciprocal_space}).

CeCoIn$_5$ crystals grow preferentially as thin plates with the $\bm{c}$ axis normal to the plate.  Our 146 mg single crystal was glued to an aluminum plate mounted on the DR to give a [1 $\bar{1}$ 0]-[0 0 1] horizontal scattering plane when the cryomagnet was level; the DR was rotated to bring the field direction close to [1 $\bar{1}$ 0].  Figure \ref{fig:reciprocal_space} shows the accessible regions of reciprocal space in this scattering plane for two orientations of the sample relative to the field.  A typical magnetic reflection close to this plane is at (1-$q$, -$q$, 1.5) = (0.56 -0.44 1.5). To bring the magnetic reflections into the scattering plane, the cryomagnet was tilted by a few degrees about the horizontal axis perpendicular to the field.  

RITA-II is a cold three axis spectrometer, with a set of seven analyzer crystals (`blades') receiving the scattered beam \cite{RITA}.  These blades reflect into a 2-D detector, and can be set up in a variety of arrangements.  We operated the instrument in its elastic scattering configuration \cite{RITA}, using the blades to probe multiple 2$\theta$ values; we assume that all scattering observed is elastic in origin.  The spectrometer and sample were always aligned such that the Bragg reflection being studied was incident on the central blade.  A beryllium filter and radial collimator were installed on the output beam.  Blade efficiency was corrected for by using incoherent scattering gathered under conditions of field or blade angle for which no magnetic signal was observed.

\begin{figure}[tbh]
\begin{center}
\includegraphics[width=0.45\textwidth]{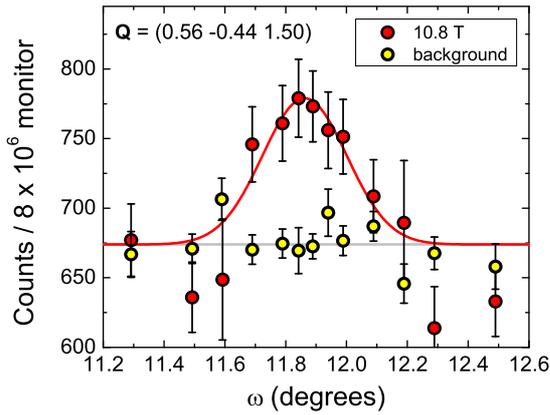}
\end{center}
\caption{(color online) Scan of the sample rotation angle $\omega$ through the magnetic peak at (0.56 -0.44 1.5), rotating the sample relative to the field and the incident neutron beam, with the magnetic field close to 12$^{\circ}$ to the basal plane.  The background is comprised of data taken at 10.9 T (all blades), and those blades showing no peak at 10.8 T.   The horizontal line is a straight line fit to the background data, and the other line is a Gaussian fit to the peak on top of the horizontal background.  The temperature of the sample was 50 mK.  A monitor count of 8000000 represents approximately two hours of counting time.}
\label{fig:a3scan}
\end{figure}

Figure \ref{fig:a3scan} shows the magnetic reflection (0.56 -0.44 1.5) measured at $T$ = 50 mK as a function of the sample rotation angle $\omega$ with a field of 10.8 T applied out of the basal plane at 11.9$^{\circ}$ to [1 $\bar{1}$ 0], known with an accuracy of $\pm$ 0.1$^{\circ}$.  The background here is a composite of data obtained at neighbouring 2$\theta$ values at 10.8 T and at 10.9 T.  The peak has been fitted as a Gaussian, giving a half-width half-maximum (HWHM) of 0.14 $\pm$ 0.04$^{\circ}$.  For nuclear Bragg reflections from the crystal, the HWHM is resolution-limited at 0.155 $\pm$ 0.001$^{\circ}$, indicating that the magnetic peak is resolution limited.  This means that the coherence length of the magnetic order is greater than 2400 \AA~along the rock direction, which is at an angle of 50$^{\circ}$ to the field.

Using all of the blades and the results for different rock angles, a scattering map can be constructed around the magnetic peak. The blades cover a region of the scattering plane around the central spot of approximately $\pm$($\delta/5$, $\delta/5$, $\delta$), where $\delta$ = 0.05 rlu.   Because of the vertical acceptance of the blades, a region out of the scattering plane of approximately $\pm$ ($\delta$,$-\delta$, 0) is also covered. The scattering map for the (1 $\bar{1}$ 0) nuclear peak gives the resolution ellipsoid, which has its longest dimension approximately parallel to the total momentum transfer $\bm{Q}$. A fit to the magnetic peak shows that within errors its shape is resolution limited.  There is no evidence in the region studied for an additional FFLO-induced modulation parallel to the field direction~\cite{YS09}, although it should be emphasized that our signal is weak, and satellites due to any modulation are expected to be weaker still.

One can evaluate the magnetic moment per Ce ion by comparing the intensity of this peak with that of the very weak (1 $\bar{1}$ 0) nuclear peak, assuming that it does not suffer from extinction.  If extinction is important, the value obtained will represent an \emph{upper limit} on the magnetic moment. Due to the geometry constraints, only two nuclear Bragg peaks were accessible, the (1 $\bar{1}$ 1) and the (1 $\bar{1}$ 0).  On comparing the intensity ratios of these two peaks to calculations of the expected intensity (including absorption corrections), there is clearly extinction of the stronger reflection, the (1 $\bar{1}$ 1). However, the extent of the extinction of (1 $\bar{1}$ 0) could not be confirmed due to the experimental constraints.  The paramagnetic contribution to the peak intensity of these nuclear peaks was not taken into account in the calculations below, but estimates indicated that it would have a negligible ($<$ 1\%) effect.

\begin{figure}[tbh]
\begin{center}
\includegraphics[width=0.45\textwidth]{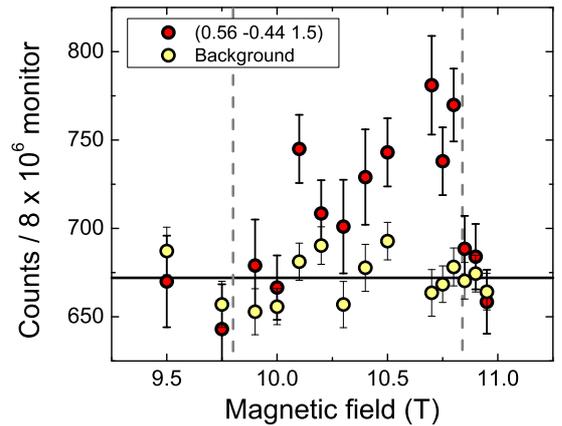}
\end{center}
\caption{(color online) The field dependence of the magnetic peak (0.56 -0.44 1.5) with a field angle of 12$^{\circ}$.  The background is the averaged sum of the data obtained from the other analyzer blades during the measurement, and so represent neighboring momentum transfer values.  Corrections for blade efficiency have been made.  The horizontal line is a straight line fit to the background points.  The vertical lines mark the transition fields as observed from magnetostriction measurements in Ref.~\onlinecite{CMMP+07}.}
\label{fig:field_dep}
\end{figure}

For the model presented in Ref.~\onlinecite{KSNS+08}, which has the modulated moments along the easy $\bm{c}$ axis, our upper limit on the Ce moment is 0.17 $\pm$ 0.02 $\mu_B$/Ce ion at 10.8 T, with the field 12$^{\circ}$ from the basal plane.  It is possible that, when the field is not in the basal plane, the modulated magnetic moments rotate to move perpendicular to the applied field.  We find that this is a small effect for a 12$^{\circ}$ tilt; the upper limit on the moment per Ce ion would become 0.16 $\pm$ 0.02 $\mu_B$.  These values are comparable with those found by Kenzelmann {\it et al.} \cite{KSNS+08,KGEG+10}, but because of the large absorption and uncertainty due to extinction, we hesitate to compare absolute magnitudes.

The field dependence of the peak was also measured (Figure \ref{fig:field_dep}), by sitting on the top of the peak and altering the field.  Limited rocking scans were carried out at 10.9 T and 9.5 T to check that the disappearance of the peak intensity was not due to its position moving with field.  The field-dependence of the peak maps very well onto the phase diagram mapped out by magnetostriction \cite{CMMP+07}, as do the results of Kenzelmann {\it et al.} \cite{KSNS+08}.

The other field angle we studied was 17$^{\circ}$, over a field range from 8.5 T to 10.5 T, and at a temperature of 50 mK. No magnetic peak was observed, probably because the magnetic moment has declined with field angle such that there is insufficient contrast.  Using the approach described above, an upper limit of 0.025 $\mu_B$ can be calculated for the modulated moment.  It is unlikely that the loss of intensity is due to a shift of the peak position, because no change in the magnetic wavevector is observed between 0$^{\circ}$ and 12$^{\circ}$. Due to the finite volume of reciprocal space surveyed, we would detect any small shift between 12$^{\circ}$ and 17$^{\circ}$.

In an antiferromagnetic structure, we expect that the modulated magnetic moments will lie along the easy axis.  If a magnetic field is applied, the modulated moments will tend to align perpendicular to the field. The magnetic structures reported in Refs.~\onlinecite{KSNS+08} and \onlinecite{KGEG+10} clearly satisfy both of these conditions: in both cases the field is in the basal plane and the moments point along the $\bm{c}$ axis.  However, when the magnetic field is rotated towards the $\bm{c}$ axis, as is done here, these conditions cannot be met simultaneously, and we expect the magnetic order to become less stable.

Our results, and the field dependent data in Ref.~\onlinecite{KSNS+08}, both match up with the phase diagram deduced from the magnetostriction measurements of Correa {\it et al.} \cite{CMMP+07}, indicating that the magnetostriction signature correlates with the existence of the magnetically ordered $Q$-phase.  This signature weakens and has disappeared when the field is applied more than 22$^{\circ}$ out of the basal plane, suggesting that this is the limiting field angle for the occurrence of the $Q$-phase.  From our limited measurements at higher angles, we find no evidence for magnetic ordering.  These two experiments together indicate that the $Q$-phase ordering gets weaker and disappears as the field is rotated out of the basal plane.  We therefore conclude that the $Q$-phase is not related to the anomalies seen in numerous bulk measurements when the field is applied parallel to the $\bm{c}$ axis; the phase associated with these anomalies must have another origin, and may be a true FFLO phase.

The origin of the magnetic order in the $Q$-phase remains an open and very interesting question.  In this phase, the ordering wavevector $\bm{q}_{mag} = (0.44,0.44,0.5)$ is effectively unchanged on rotating the applied field direction - both in and out of the basal plane, and is close to that observed when superconductivity is suppressed by Cd-doping \cite{NSPH+07}. A spin resonance peak is seen at $\bm{q}_{SR} = (\frac{1}{2} \frac{1}{2} \frac{1}{2})$ \cite{SBHK+08}, and moves to lower energy as a field is applied \cite{PRLF09}. These all point towards the ordering being controlled by a peak in susceptibility at a characteristic Fermi surface nesting vector \cite{Machida}, presumably between the dominant quasicylindrical sheets running parallel to $\bm{c}^*$ \cite{KOEB+09}. Nonetheless, the appearance of the $Q$-phase only within the superconducting state has to be explained. The repeat of the magnetic structure in the basal plane is $a / (\sqrt(2) q) \sim$ 7 \AA.  The distance between planes of minimum Ce moment in the $Q$-phase $=a/(\sqrt{2}\times 2(0.5-q))\sim$ 27 \AA.  Both of these lengths are considerably smaller than the superconducting coherence length, indicating that the dominant order parameter is probably not spatially-varying triplet superconductivity \cite{AVLS08,AST09}.  Instead, the disappearance of the $Q$-phase in the normal state, and its suppression when the field is moved out of the basal plane suggests that the magnetic order just becomes stable because of a small enhancement of the antiferromagnetic susceptibility in the mixed state \cite{Machida}, that may be brought about by strong Pauli paramagnetic depairing \cite{IHA10}, or is perhaps driven by an FFLO-type pair density wave \cite{YS09,KSMH+10}, which would survive the magnetic ordering as the field direction is moved towards $\bm{c}$.  The confirmation of this model requires greater intensity to observe satellite magnetic peaks expected parallel to the applied field, or NMR measurements as a function of field angle. 

This work is based on experiments performed at the Swiss spallation neutron source SINQ,  Paul Scherrer Institute, Villigen, Switzerland.  EB and EMF acknowledge support from the U.~K.~EPSRC and the European Commission under the 7th Framework Programme through the `Research Infrastructures' action of the `Capacities' Programme, Contract No CP-CSA-INFRA-2008-1.1.1 Number 226507-NMI3. PD and MRE acknowledge financial support by the U.~S.~NSF through grant DMR-0804887.  ML acknowledges support from DanScatt.  Work carried out at Brookhaven National Laboratory (CP) was supported by the U.S. Department of Energy and Brookhaven Science Associates (No. DE-Ac02-98CH10886).  JSW acknowledges financial support from the Swiss National Centre of Competence in Research program `MaNEP'.

\end{document}